# Anomalous Compressibility Effects and Superconductivity in $EuFe_2As_2$ under High Pressures

**Walter Uhoya, Georgiy Tsoi and Yogesh K. Vohra**
Department of Physics, University of Alabama at Birmingham (UAB)
Birmingham, AL 35294, USA

**Michael A. McGuire, Athena S. Sefat, Brian C. Sales, and David Mandrus**
Oak Ridge National Laboratory (ORNL)
Oak Ridge, TN 37831, USA

**Samuel T. Weir**
Mail Stop L-041, Lawrence Livermore National Laboratory (LLNL)
Livermore, CA 94550, USA

The crystal structure and electrical resistance of the structurally-layered $EuFe_2As_2$ have been studied up to 70 GPa and down to temperature of 10 K, using a synchrotron x-ray source and the designer diamond anvils. The room-temperature compression of the tetragonal phase of $EuFe_2As_2$ (*I4/mmm*) results in an increase in the *a*-axis and a rapid decrease in *c*-axis with increasing pressure. This anomalous compression reaches a maximum at 8 GPa and the tetragonal lattice behaves normal above 10 GPa with a nearly constant *c/a* axial ratio. The rapid rise in superconducting transition temperature ($T_c$) to 41 K with increasing pressure is correlated to this anomalous compression and a decrease in $T_c$ is observed above 10 GPa. We present P-V data or equation of state of $EuFe_2As_2$ in both the ambient tetragonal phase and the high pressure collapsed tetragonal phase to 70 GPa.

The discovery of a new class of iron arsenide superconductors [1] has provided fresh impetus to high pressure studies on these materials whereby pressure variable provides a controlled tuning of superconducting properties without the disorder effects induced by chemical substitution or doping. The so called 122 iron-based superconducting materials $AFe_2As_2$ (A = divalent alkali earth or rare earth metal Ca, Sr, Ba, and Eu) are of particular interest because of the overlap of superconducting and anti-ferromagnetic phases in these materials. In particular, 122 compound $EuFe_2As_2$ has been investigated to only modest pressures of 3 GPa and pressure-induced superconductivity has been reported beginning at approximately 2 GPa [2, 3]. The high pressure studies on $EuFe_2As_2$ are intriguing because a simultaneous occurrence of superconductivity and antiferromagnetic ordering is observed at high pressures and low temperatures [2, 3]. The structural and magnetic phase transitions have been extensively investigated at ambient pressure and at low temperatures in $EuFe_2As_2$ [4, 5] and reveal two phase transitions at 190 K and 20 K respectively. The phase transition $T_0 = 190$ K at ambient pressure is attributed to a Spin Density Wave (SDW) and results in a crystallographic phase transition from ambient tetragonal phase (*I4/mmm*) to an orthorhombic phase (*Fmmm*) and the Fe moments order antiferromagnetically. The phase transition at 20 K is an antiferromagnetic ordering of the $Eu^{2+}$ moments and is referred to as the Nèel temperature ($T_N$). The SDW transition is suppressed by the application of high pressure and transition temperature $T_0$ is found to decrease rapidly with increasing pressure and SDW transition is not detected above 2.5 GPa. The previous high pressure studies on $EuFe_2As_2$ indicate that $T_N$ is relatively insensitive to pressure and remains at 20 K up to 3 GPa, while superconductivity transition temperature $T_c$ appears at 30 K at a pressure of 2.8 GPa [2, 3].

The present study is motivated by the occurrence of superconductivity under high pressure in $EuFe_2As_2$ and its correlation to the compression of the tetragonal (T) lattice and the possible occurrence of collapsed tetragonal (CT) phase under high pressures. The compression behavior of $BaFe_2As_2$ has been studied up to 22 GPa at ambient temperature and anisotropic compression effects have been documented [6]. The first-and second-order phase transitions in ternary europium phosphides with $ThCr_2Si_2$-type structure have been reported under high pressure [7]. Also, a non-magnetic CT phase has been documented in $CaFe_2As_2$ at P= 0.24 GPa and T = 50 K [8] and the impact of CT phase on superconducting properties of $CaFe_2As_s$ has been extensively studied [9-11]. An additional consideration for present studies is that the high-pressure Mössbauer investigations on $EuFe_2P_2$ have revealed a continuous valence transition from a magnetic $Eu^{2+}$ state to a nonmagnetic $Eu^{3+}$ state in the pressure range between 3-9 GPa [12]. We have carried out detailed structural studies on $EuFe_2As_2$ under ultra high pressures to 70 GPa at ambient temperature using image plate x-ray diffraction at a synchrotron source.

The single crystal samples of $EuFe_2As_2$ were grown based on FeAs self-flux methods as described in reference [13]. The single crystal sample was powdered and loaded in a diamond anvil cell for high pressure x-ray diffraction experiments. Our x-ray diffraction studies at ambient pressure and temperature revealed a tetragonal structure with lattice parameters $a = 3.916 \pm 0.012$ Å and $c = 12.052 \pm 0.036$ Å with axial ratio $c/a = 3.078$ at ambient temperature and pressure. The tetragonal crystal structure of $EuFe_2As_2$ is identified as $ThCr_2Si_2$ type with space group *I4/mmm* with Eu atoms at 2a position (0, 0, 0), Fe atoms at 4d positions (0, ½, 1/4) and (1/2, 0, ¼), As atoms at 4e position (0, 0, z) and (0, 0, -z). The structural parameter $z = 0.3625$ has been obtained from Rietveld refinement of x-ray diffraction data [4]. The high pressure x-ray diffraction experiments were carried out at the beam-line 16-ID-D, HPCAT, Advanced Photon Source, Argonne National Laboratory. An angle dispersive technique with an image-plate area detector was employed using a x-ray wavelength $\lambda = 0.4072$ Å. We employed eight probe designer diamond anvils [14, 15] in high pressure four-probe electrical resistance measurements on the $EuFe_2As_2$ compound. The eight tungsten microprobes are encapsulated in a homoepitaxial diamond film and are exposed only near the tip of the diamond to make contact with the $EuFe_2As_2$ sample at high pressure. Two electrical leads are used to set constant current through the sample and the two additional leads are used to monitor the voltage across the sample. The pressure was monitored by the ruby fluorescence technique and care was taken to carefully calibrate ruby $R_1$ emission to low temperature of 10 K as described in an earlier publication [16]. In x-ray diffraction experiments, an internal copper pressure standard was employed for the calibration of pressure [17]. The Birch Murnaghan equation [18] as shown by equation (1) was fitted to the available equation of state data on copper pressure standard [17].

$$P = 3B_0 f_E (1+2f_E)^{5/2} \left\{1+\frac{3}{2}(B'-4)f_E\right\} \qquad (1)$$

Where $B_o$ is the bulk modulus, $B'$ is the first derivative of bulk modulus at ambient pressure, and $V_0$ is the ambient pressure volume. The fitted values for copper pressure standard are $B_0$ = 121.6 GPa, $B'$ = 5.583, and $V_0$ = 11.802 Å$^3$/atom. The parameter $f_E$ is related to the volume compression and is described below:

$$f_E = \frac{\left[\left(\frac{V_o}{V}\right)^{2/3}-1\right]}{2} \qquad (2)$$

Fig. 1 shows the integrated x-ray diffraction profiles for EuFe$_2$As$_2$ obtained at various pressures from the image plate diffraction studies at the beam-line 16-ID-B, Advanced Photon Source, Argonne National Laboratory using a wavelength $\lambda$ = 0.4072 Å. The image plate x-ray diffraction patterns were recorded with a 5 microns x 5 microns focused x-ray beam on an 80 microns diameter EuFe$_2$As$_2$ sample mixed with copper pressure marker. The use of a focused x-ray beam allows us to collect high quality x-ray diffraction patterns that are not affected by the inhomogeneous pressure conditions over the sample. The geometrical constraints in our diamond anvil cell device allowed x-ray diffraction data to be obtained to 2θ below 15 degrees (or interplanar $d$-spacing > 1.56 Å). Fig. 1 (a) shows the x-ray diffraction pattern at 0. 4 GPa with EuFe$_2$As$_2$ sample in the *I4/mmm* tetragonal phase and copper pressure marker in the face centered cubic (*fcc*) phase. The diffraction peaks are labeled by their respective (hkl) values. The tetragonal phase of EuFe$_2$As$_2$ is characterized by the (101), (110), (112), (200), (211), (204), and (213) diffraction peaks. The *fcc* phase of copper pressure marker is characterized by the (111) and (200) diffraction peaks. The measured volume of copper by x-ray diffraction was used to calculate sample pressure from the equation of state given by equation (1). The measured lattice parameters at 0.4 GPa are $a$ = 3.923 ± 0.012 Å and $c$ = 11.985 ± 0.036 Å. Fig. 1 (b) shows the x-ray diffraction spectrum at a pressure of 8.5 GPa with lattice parameters $a$ = 3.989 ± 0.012 Å and $c$ = 10.006 ± 0.030 Å. The EuFe$_2$As$_2$ (101), (112), and (213) diffraction peaks clearly show a movement to higher diffraction angles or lower interplanar spacing with increasing pressure while the (110) and (200) move to lower diffraction angles or higher interplanar spacing when comparing x-ray diffraction patterns at 8.5 GPa to 0.4 GPa (Fig. 1(a) and Fig. 1(b)). This is a clear evidence of the anomalous compression of the tetragonal lattice as the $a$-axis increases with increasing pressure and the $c$-axis decreases with increasing pressure. This is in contrast to normal compression behavior which shows a uniform decrease in all dimensions of the unit cell. Another important point to note is that EuFe$_2$As$_2$ (213) diffraction peak remains a single peak without any evidence of splitting and hence there is no evidence of a tetragonal to orthorhombic phase transformation at high pressures as has been noted at low temperatures [4, 5]. Fig. 1 (c) shows that the phase remains tetragonal until the highest pressure of 70.2 GPa with lattice parameters $a = 3.692 \pm 0.011$ Å and $c = 9.059 \pm 0.027$ Å. The compression of the tetragonal lattice between 8.5 GPa and 70.2 GPa can be considered normal as all diffraction peaks move to higher angles or lower interplanar $d$-spacings as can be seen by comparing Fig. 1(b) and Fig. 1(c).

Fig. 2 shows the measured lattice parameters *"a"* and *"c"* as a function of pressure exhibiting anomalous compression effects. The lattice parameter *"a"* shows an increase with increasing pressure with a peak at a pressure of 8.5 GPa. A further increase in pressure beyond 8 GPa, results in a decrease in *"a"* and a normal compression behavior that continues until the highest pressure of 70 GPa. This negative compressibility of *"a"* lattice parameter is a very intriguing feature of EuFe$_2$As$_2$ compression. The lattice parameter *"c"* shows a rapid decrease with increasing pressure until 8 GPa and a normal decrease with further increase in pressure to 70 GPa.

Fig. 3 shows the measured axial ratio (*c/a*) as a function of pressure to 70 GPa at ambient temperature. The axial ratio (*c/a*) shows a rapid decrease with increasing pressure till 8 GPa and a gradual decrease above this pressure. In fact, the *c/a* ratio variation as a function of pressure can be divided in to two linear regions and the fits for the two linear regions are shown in Fig. 3 and described below.

$$c/a = 3.033 - 0.075 \, P, \qquad 0 \leq P \leq 8 \text{ GPa} \qquad (3)$$

$$c/a = 2.536 - 0.001\,P, \qquad 10 \leq P \leq 70 \text{ GPa} \qquad (4)$$

The intersection of the two linear regions as described by equations (3) and (4) defines the phase transition from the ambient pressure T-phase to the so called CT-phase. This phase transition is defined by the present experiments to occur near 8 GPa at ambient temperature. A further x-ray diffraction study at low temperature would be needed to follow this phase boundary to low temperatures.

Fig. 4 shows the measured Pressure-Volume (P-V) curve or equation of state for $EuFe_2As_2$ to 70 GPa at ambient temperature. Also shown are the Birch Murnaghan equation of state fits as described by equation (1). The fit parameters for the tetragonal and collapsed tetragonal phases are summarized in Table 1. The fitted ambient pressure volume ($V_0$) is 184.5 Å$^3$ for the T-phase and is 170.0 Å$^3$ for the CT-phase indicating that a hypothetical CT-phase at ambient pressure has a density that is 7.9 % higher than the T-phase (Table 1). The measured equation of state shows considerable stiffening at the tetragonal to collapsed tetragonal phase transition as evidenced by an abrupt change in slope of the volume-pressure curve at 8 GPa as shown in Fig. 4. The fitted value of Bulk Modulus $B_0 = 39.3$ GPa for the T-phase and $B_0 = 134.0$ GPa for the CT-phase at ambient pressure. This comparison shows that the T-phase of $EuFe_2As_2$ is more compressible than a hypothetical CT phase at ambient pressure by a factor of 3.4.

Fig. 5 shows the measured superconducting transition temperature $T_c$ as a function of pressure. The $T_c$ measurements have been carried out using four probe electrical resistance measurements using designer diamond anvil as described in an earlier publication [16]. The insert in Fig. 5 shows electrical resistance as a function of temperature at a pressure of 10.3 GPa indicating onset of superconductivity $T_c$ at 41 K. It is interesting to note that $T_c$ shows a rapid increase in the pressure region where anomalous compression effects were observed in x-ray diffraction experiments. In particular, $T_c$ shows a rapid increase between 4 GPa and 10 GPa and attains a value as high as 41 K near 10 GPa. A further increase in pressure beyond 10 GPa leads to reduction in $T_c$ in the CT phase. It seems that anomalous compression tends to favor superconductivity as the FeAs tetrahedral bonded layers are brought together and then shows a decrease as we reach the compression limit of the tetragonal phase and as a collapsed tetragonal structure is formed. The structural distortions under pressure and chemical doping in superconducting $BaFe_2As_2$ and $CeFeAsO_{1-x}F_x$ have been correlated to the superconducting properties [19, 20]. In particular, superconducting transition temperature $T_c$ is found to increase in $BaFe_2As_2$ as As-Fe-As bond angles tend to the ideal tetrahedral value of 109.5 deg [19]. Our measured lattice parameters $a = 3.989 \pm 0.012$ Å and $c = 10.006 \pm 0.030$ Å at 8.5 GPa for $EuFe_2As_2$ along with an estimated structural parameter $z = 0.370$ results in As-Fe-As bond angles of 105.5 deg and 117.7 deg. The structural parameter $z$ is not well constrained in our experiments due to limited diffraction data set and preferred alignment of crystal grains common in high pressure experiments. The estimated As-Fe-As bond angle values for $EuFe_2As_2$ are far removed from the ideal bond angles of 109.5 deg observed in $BaFe_2As_2$, however, it should be noted that $EuFe_2As_2$ is different from the other 122 superconducting materials due to a valence transition from a magnetic $Eu^{2+}$ state to a nonmagnetic $Eu^{3+}$ state that occurs in the pressure range between 3 GPa and 9 GPa [12]. A complete description of the superconducting properties of $EuFe_2As_2$ would likely involve both the structural changes documented in this paper and the valence transition recorded in high pressure Mössbauer investigations.

In summary, we have studied 122 iron-based layered compounds $EuFe_2As_2$ to 70 GPa at ambient temperature using a synchrotron source. The image plate x-ray diffraction studies reveal anomalous compressibility effects whereby *a*-axis of the tetragonal phase shows an increase with increasing pressure while the *c*-axis shows a rapid decrease with increasing pressure. A maximum in *a*-axis is observed at a pressure of 8 GPa and a normal compression behavior is observed above this pressure. The phase above 8 GPa is referred to as the collapsed tetragonal (CT) phase as opposed to the normal tetragonal (T) phase. The equation of state for the T-phase and CT-phases show distinct Bulk Modulus ($B_0$) and pressure derivative of Bulk Modulus ($B´$). At ambient pressure, an extrapolated CT phase has a density that is 7.9 % higher as compared to the T-phase under similar conditions. The superconductivity in $EuFe_2As_2$ has been studied under high pressure using four probe electrical resistance measurements with designer diamond anvils. The superconductivity transition temperature ($T_c$) is marked by a decrease in electrical resistance as a function of temperature. The measured value of $T_c$ shows an increase between 4 GPa and 10 GPa and attains a value as high as 41 K under high pressure. This increase in $T_c$ is correlated with the anomalous compressibility effects documented in this report. The superconducting transition $T_c$ shows a gradual decrease in pressure above 10 GPa in the CT-phase. Our high pressure studies on 122 iron-based layered superconductor $EuFe_2As_2$ have shown anomalous compressibility effects, presence of a collapsed tetragonal phase, and superconducting properties that are strongly linked to its compression behavior.

## ACKNOWLEDGMENT

Walter Uhoya acknowledges support from the Carnegie/Department of Energy (DOE) Alliance Center (CDAC) under Grant No. DE-FC52-08NA28554. Research at ORNL is sponsored by the Division of Materials Sciences and Engineering, Office of Basic Energy Sciences, U.S. Department of Energy. Portions of this work were performed at HPCAT (Sector 16), Advanced Photon Source (APS), Argonne National Laboratory.

**Figure Captions:**

Fig. 1: The integrated x-ray diffraction profiles for EuFe$_2$As$_2$ at various pressures at ambient temperature using a x-ray wavelength λ = 0.4072 Å. (a) Diffraction pattern at a low pressure of 0.4 GPa. (b) Diffraction pattern at 8.5 GPa showing anomalous compression effects discussed in the text. (c) Diffraction pattern at the highest pressure of 70.2 GPa showing stability of the tetragonal phase. The copper (Cu) pressure marker diffraction peaks are present in all spectra.

Fig. 2: The measured lattice parameters *"a"* and *"c"* for the tetragonal phase of EuFe$_2$As$_2$ as a function of pressure. A negative compressibility is observed for the *a*-axis and a maximum is observed at 8.5 GPa. The *c*-axis shows a rapid decrease with increasing pressure till 8 GPa and a normal decrease with further increase in pressure. The error bars for *"c"* are less than the symbol size used in plotting.

Fig. 3: The measured axial ratio (*c/a*) for the tetragonal phase of EuFe$_2$As$_2$ as a function of pressure to 70 GPa. The linear fits for the two phases, i.e., the Tetragonal (T) phase and the Collapsed Tetragonal (CT) phase are described in the text. The transition between the two phases occurs at 8 GPa at ambient temperature.

Fig. 4: The measured equation of state data for the tetragonal phase of EuFe$_2$As$_2$ as a function of pressure to 70 GPa. The solid curves are the Birch Murnaghan equation of state fit to the two phases, i.e., the Tetragonal (T) phase and the Collapsed Tetragonal (CT) phase. The fit parameters are summarized in Table 1. The transition between the two phases occurs at 8 GPa at ambient temperature.

Fig. 5: The measured superconducting transition (T$_c$) temperature as a function of pressure for EuFe$_2$As$_2$ as obtained by electrical resistivity measurements. T$_c$ shows an increase between 4 GPa and 10 GPa and gets to as high as 41 K. The T$_c$ shows a gradual decrease above 10 GPa and correlates with the anomalous compressibility effects in this material. The insert shows electrical resistance as a function of temperature at 10.3 GPa indicating onset of superconductivity at 41 K.

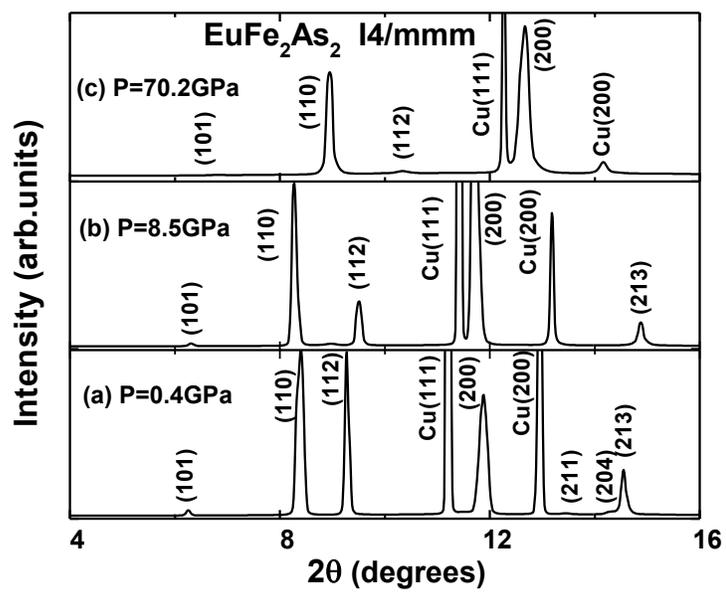

Figure 1

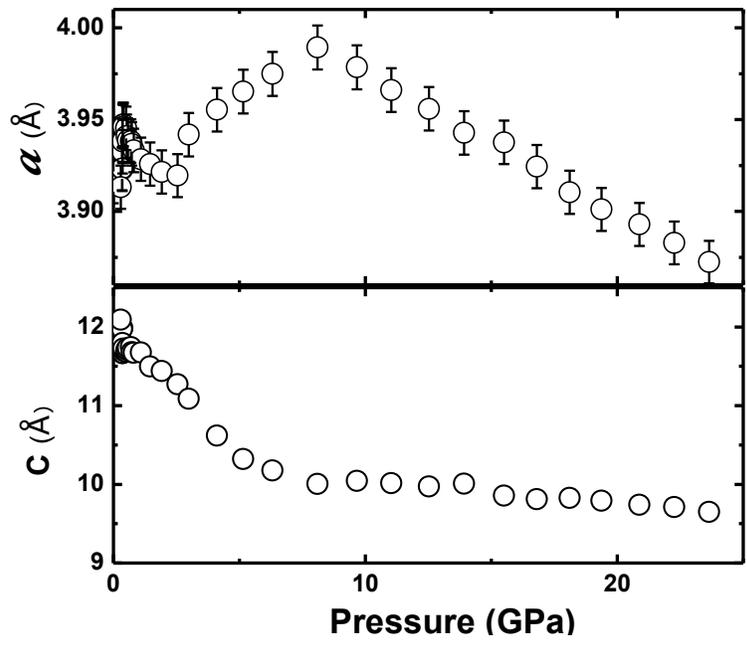

Figure 2

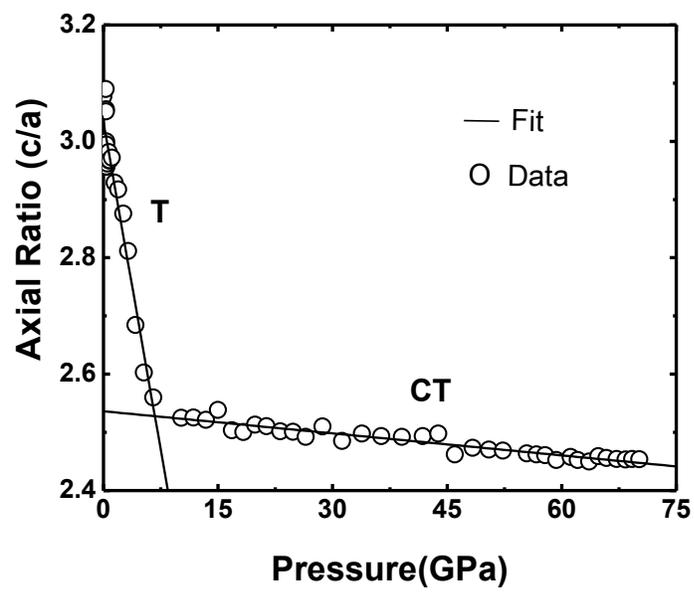

Figure 3

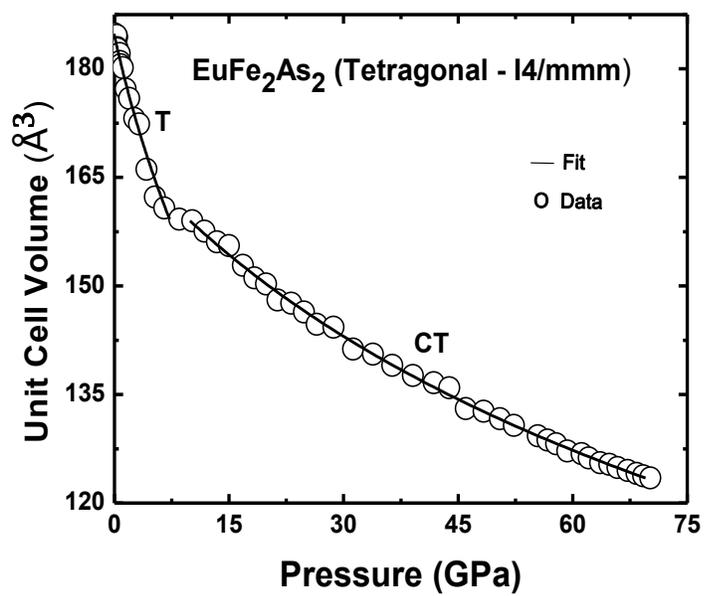

Figure 4

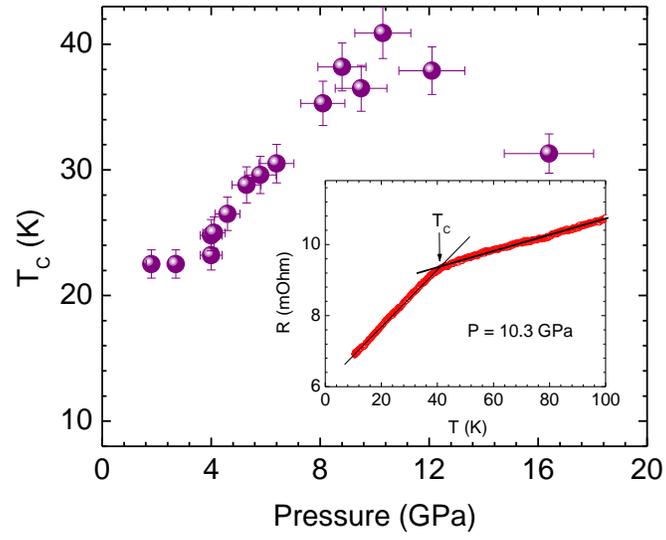

Figure 5

Table 1: Equation of state parameters for the EuFe$_2$As$_2$ sample in the Tetragonal (T) phase and Collapsed Tetragonal (CT) phase at ambient temperature.

| Phase | Unit Cell Volume ($V_0$) Ambient Conditions | Bulk Modulus ($B_0$) | Pressure Derivative of Bulk Modulus ($B'$) |
|---|---|---|---|
| Tetragonal (T) 0 < P < 8 GPa | 184.5 ± 0.4 Å$^3$ | 39.3 ± 1.6 GPa | 6.9 ± 0.4 |
| Collapsed Tetragonal (CT) 10 GPa < P < 70 GPa | 170.0 ± 0.4 Å$^3$ | 134.0 ± 1.6 GPa | 3.3 ± 0.2 |